\documentclass{aa}
\usepackage{graphicx}
\usepackage{txfonts}
\usepackage{natbib}
\bibpunct{(}{)}{;}{a}{}{,}
\usepackage{longtable}
\usepackage{lscape}
\usepackage{times}
\usepackage{textcomp}

\begin{document}

\title{The Origem Loop}
\subtitle{}
\author{X. Y.~Gao and J. L.~Han}
\titlerunning{The Origem Loop}
\authorrunning{Gao \& Han}

\offprints{bearwards@gmail.com}

\institute{National Astronomical Observatories, Chinese Academy of
  Sciences, Jia-20 Datun Road, Chaoyang District, Beijing 100012, PR
  China}

\date{Received; accepted}

\abstract
{The Origem Loop in the Galactic anticentre was discovered in 1970s
  and suggested to be a large supernova remnant. It was argued later
  to be a chance superposition of unrelated radio sources.}
{We attempt to understand the properties of the Origem Loop.}
{Available multi-frequency radio data were used for the determination
  of radio spectra of different parts of the Origem Loop and the
  polarization properties of the loop.}
{Newly available sensitive observations show that the Origem Loop is a
  loop of more than 6$\degr$ in diameter. It consists of a large
  non-thermal arc in the north, which we call the Origem Arc, and
  several known thermal \ion{H}{II} regions in the south.  Polarized
  radio emission associated with the arc was detected at
  $\lambda$6\ cm, revealing tangential magnetic fields. The arc has a
  brightness temperature spectral index of $\beta = -2.70$, indicating
  its non-thermal nature as a supernova remnant. We estimate the
  distance to the Origem Arc to be about 1.7~kpc, similar to those of
  some \ion{H}{II} regions in the southern part of the loop.}
{The Origem Loop is a visible loop in the sky, which consists of a
  supernova remnant arc in the north and \ion{H}{II} regions in the
  south.}


\maketitle
\section{Introduction}

Several giant loops were recognized in the early radio sky maps, i.e.
Loop I \citep{Brown60}, Loop II \citep{Large62}, Loop III
\citep{Quigley65}, and Loop IV \citep{Large66}. By comparing the radio
continuum, H$\alpha$, interstellar polarization and the \ion{H}{I}
observations of the four loops, \citet{Haslam71} made a general review
of these giant structures.  \citet{Berkhuijsen71} summarized the
geometric parameters of the four loops, and proposed their origin from
supernova explosions.  Besides these four giant loops, there are also
loops with smaller sizes, i.e. the Lupus Loop \citep{Gardner65} and
the Cygnus Loop \citep{Walsh55} which have been undoubtedly identified
as supernova remnants (SNR) and are collected in the well-known SNR
catalog compiled by Dave Green \citep{Green09}.

The Origem Loop is another known Galactic radio loop discovered in
1970's by \citet{Berkhuijsen74} on the 178~MHz radio map
\citep{Caswell69} in the Galactic anti-centre region between the
constellations Orion and Gemini. However, its nature is under
debate. The loop was first proposed to be an old SNR at a distance of
about 1~kpc with a diameter of about 5$\degr$ \citep{Berkhuijsen74},
but was later argued by \citet{Caswell85} to be a possible projection
effect of several unrelated \ion{H}{II} regions, many extra-Galactic
sources and a discrete small SNR G192.8$-$1.1 (PKS 0607+17) with a
diameter of about 80$\arcmin$. Note, however, that \citet{Gao11x} have
disproved G192.8$-$1.1 to be a SNR but a thermal emitter using the
Urumqi $\lambda$6\ cm \citep{Gao10}, the Effelsberg $\lambda$11\ cm
\citep{Fuerst90} and the Effelsberg $\lambda$21\ cm \citep{Reich97}
survey data.  Probably because of its large size, there were few
follow-up observations of the Origem Loop after
\citet{Berkhuijsen74}. \citet{Caswell85} discussed the region of
G192.8$-$1.1 and some nearby \ion{H}{II} regions, but did not study
the northern part of the Origem Loop. \citet{Krymkin88} made
brightness temperature scans to nearly the entire loop with the UTR-2
and RATAN 600 radio telescopes at five frequencies, from 14.7~MHz to
3950~MHz.  They claimed the discovery of a new feature, namely GR
0625+16, to be another possible discrete SNR besides the ``SNR''
G192.8$-$1.1. However, the GR 0625+16 exactly corresponds to the
northern arc of the Origem Loop.

High quality multi-frequency radio survey data with sufficient
sensitivity and angular resolution are now available, which can be
used to investigate the properties of the Origem Loop.  We introduce
in Sect.~2 the data sets we use, and present the analysis in
Sect.~3. A summary is given in Sect.~4.

\begin{figure*}
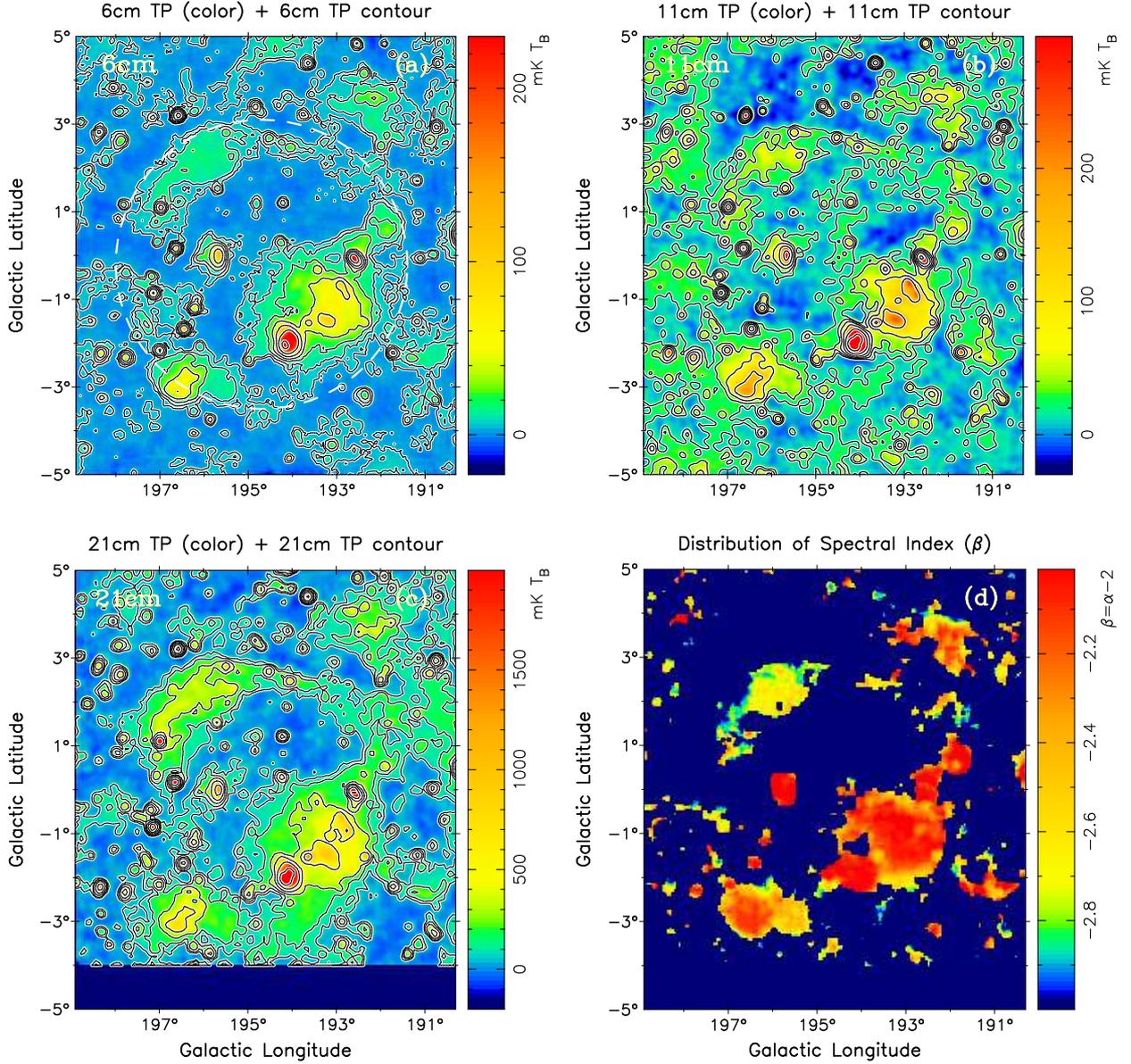
 
\centering
\begin{minipage}[bth]{0.45\textwidth}
\centering
\includegraphics[angle=-90, width=8.0cm]{Gshell_6i_color.ps}
\end{minipage}
\begin{minipage}[bth]{0.45\textwidth}
\centering
\includegraphics[angle=-90, width=8.0cm]{Gshell_11i_color.ps}
\end{minipage}
\begin{minipage}[bth]{0.45\textwidth}
\centering
\vspace{0.5cm}
\includegraphics[angle=-90, width=8.0cm]{Gshell_21i_color.ps}
\end{minipage}
\begin{minipage}[bth]{0.45\textwidth}
\centering
\vspace{0.5cm}
\includegraphics[angle=-90, width=8.0cm]{spectral_map_3sigma.ps}
\end{minipage}
\caption{From {\it top left (a), top right (b)} to {\it bottom left
    panel (c)}: $\lambda$6\ cm, $\lambda$11\ cm, and the
  $\lambda$21\ cm total intensity images of the Origem Loop. The
  angular resolutions are 9$\farcm$5, 9$\farcm$5, and 9$\farcm$4,
  respectively.  The contours run from $2^{n}\times3.6~(3\sigma)~{\rm
    mK}\ T_{B} $, (n = 0, 1, 2 ...) for the $\lambda$6\ cm image,
  $2^{n}\times18.0~(3\sigma)~{\rm mK}\ T_{B} $, (n = 0, 1, 2 ...) for
  the $\lambda$11\ cm image, and $2^{n}\times66.0~(3\sigma)~{\rm
    mK}\ T_{B} $, (n = 0, 1, 2 ...) for the $\lambda$21\ cm image. The
  white circle in the {\it top left panel (a)} indicates the boundary
  of the Origem Loop. {\it Bottom right panel (d):} Spectral index
  distribution for the Origem Loop area derived from the Urumqi
  $\lambda$6\ cm, Effelsberg $\lambda$11\ cm, and $\lambda$21\ cm
  images at the same angular resolution of 9$\farcm$5.}
\label{Gshell}
\end{figure*}

\section{Data}

The Origem Loop clearly shows up in the $\lambda$6\ cm total intensity
and polarization images from the Sino-German $\lambda$6\ cm
polarization survey of the Galactic
plane\footnote{http://zmtt.bao.ac.cn/6cm/}\citep{Gao10}, which
motivated us to seek for a better understanding of this large
structure. Other public radio data are available from the Effelsberg
$\lambda$11\ cm \citep[2.7~GHz,][]{Fuerst90} and $\lambda$21\ cm
(1.4~GHz) Galactic plane survey \citep{Reich97}, which can be
downloaded from the survey sampler of the Max-Planck-Institut f{\"u}r
Radioastronomie
(MPIfR)\footnote{http://www.mpifr.de/old$_{-}$mpifr/survey.html}, the
WMAP 7-year K-band (22.8~GHz, $\lambda$1.3\ cm) survey data
\citep{Jarosik11} retrieved from the website of
NASA\footnote{http://lambda.gsfc.nasa.gov/product/map/dr4/maps$_{-}$band$_{-}$r9$_{-}$iqus\\$_{-}$7yr$_{-}$get.cfm},
the $\lambda$21\ cm Effelsberg Medium Latitude Survey (EMLS) data
\citep{Uyaniker98, Uyaniker99}, and the DRAO $\lambda$21\ cm
polarization survey data \citep{Wolleben06}, both of which were also
obtained from the survey sampler of MPIfR. In the following
discussions, we used the observing wavelength to indicate the data
set. Specifically the $\lambda$21\ cm data stands for the Effelsberg
Galactic plane survey data \citep{Reich97}, unless special statements
are made. The angular resolution is 9$\farcm$5 for the $\lambda$6\ cm
image, 4$\farcm$3 for $\lambda$11\ cm image, 9$\farcm$4 for
$\lambda$21\ cm image, 52$\farcm$8 for $\lambda$1.3\ cm, 9$\farcm$35
for EMLS $\lambda$21\ cm, and 36$\arcmin$ for DRAO $\lambda$21\ cm
images. Among them, the $\lambda$6\ cm and $\lambda$1.3\ cm
observations provided both total intensity and polarization
measurements, the DRAO $\lambda$21\ cm data provided only the
polarization image, while the rest give only the total intensity
maps. The Effelsberg $\lambda$21\ cm Galactic plane survey
\citep{Reich97} has a latitude limit of $b = \pm4\degr$. Therefore, we
used the EMLS data to fill the blank region above $b = 4\degr$, but we
do not have data for the region below $b = -4\degr$. Basic parameters
of these data sets are summarized in Table~1. As done by
\citet{Gao11y}, the ``background filter'' technique developed by
\citet{Sofue79} was applied to $\lambda$6\ cm, $\lambda$11\ cm and
$\lambda$21\ cm images to separate the unrelated large-scale Galactic
emission from the Origem Loop emission. A twisted hyper plane defined
by the corner mean values of each image was subtracted to find the
local zero level around the Origem Loop. The final $\lambda$6\ cm,
$\lambda$11\ cm, $\lambda$21\ cm total intensity images of the Origem
Loop are shown in Fig.~\ref{Gshell}. The $\lambda$11\ cm image was
convolved to an angular resolution of 9$\farcm$5 to get a higher
signal-to-noise ratio.

\begin{table*}
\centering
   \caption{Parameters of the survey data for the images of the Origem
     Loop}
   \label{t1}
  {\begin{tabular}{rcccr}\hline\hline

\multicolumn{1}{c}{Surveys}     &\multicolumn{1}{c}{Frequency}           &\multicolumn{1}{c}{HPBW}         &\multicolumn{1}{c}{r.m.s}     &\multicolumn{1}{c}{References} \\
                                &\multicolumn{1}{c}{(GHz)}                 &\multicolumn{1}{c}{($\arcmin$)}    &\multicolumn{1}{c}{(mK T$_{b}$)}  &\\
\hline

 Urumqi $\lambda$6\ cm          &4.8                                     &9.5                              &1.2(TP)/0.5(PI)     & \citet{Gao10}\\
 Effelsberg $\lambda$11\ cm     &2.7                                     &4.3                              &18.0(TP)            & \citet{Fuerst90}\\ 
 Effelsberg $\lambda$21\ cm     &1.4                                     &9.4                              &22.0(TP)            & \citet{Reich97}\\
 WMAP $\lambda$1.3\ cm          &22.8                                    &52.8                             &0.07(TP)/0.06(PI)   & \citet{Jarosik11}\\ 
 DRAO $\lambda$21\ cm           &1.4                                     &36.0                             &12.0(PI)            & \citet{Wolleben06}\\
 EMLS $\lambda$21\ cm           &1.4                                     &9.35                             &15.0(TP)            & \citet{Uyaniker98}\\
 \hline
\multicolumn{5}{l}{{\bf Note:} TP: total power, PI: polarization intensity}
\end{tabular}}
\end{table*}

\section{Results}

\begin{figure} 
\centering
\includegraphics[angle=-90, width=0.45\textwidth]{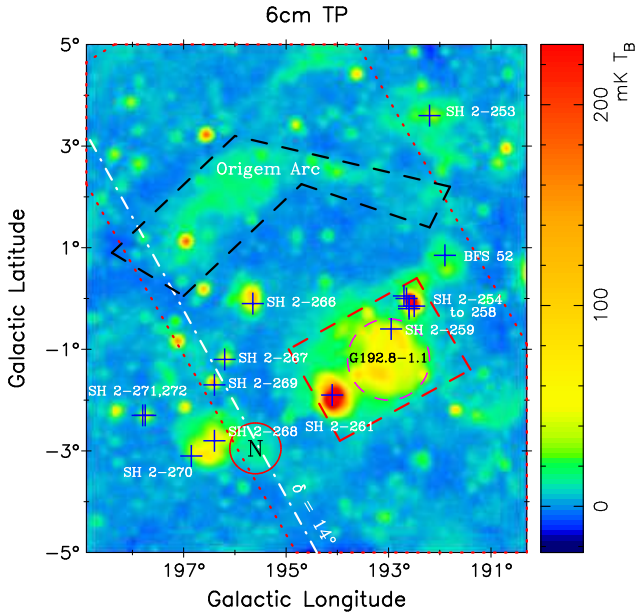}
\caption{$\lambda$6\ cm total intensity image with the prominent
  \ion{H}{II} regions marked with ``$+$'' and labeled with the
  names. The disproved SNR, G192.8$-$1.1, was also marked with a
  circle of the pink dashed line. The outer red dotted line delineates
  the common area of the observations by \citet{Krymkin88} and our
  image, while the red dashed line indicates the field shown in
  \citet{Caswell85}. The white dashed-dot line shows the declination
  of $\delta = 14\degr$ (Epoch 1950). The region on the left side of
  this line was not included in the 178~MHz map used by
  \citet{Berkhuijsen74}. The area outlined by the black dashed line,
  containing the Origem Arc, and was used for the TT-plot in
  Sect.3.1. A probable new \ion{H}{II} region G195.60-2.95 is marked
  using a circle with the letter ``N'' inside.}
\label{name}
\end{figure}

Radio images in Fig.~\ref{Gshell} at three different wavelengths
resemble each other in structures. At $\lambda$6\ cm, the circle which
indicates the loop in Fig~\ref{Gshell} (a) has a radius of
200$\arcmin$ and is centred at $\ell = 194\fdg7, b=-0\fdg2$. These
values are different from those of \citet{Berkhuijsen74}, because we
included the region below $\delta = 14\degr$ (B1950), which was not
included in the 178~MHz map used by \citet{Berkhuijsen74}. Our new
sensitive measurements enable us to detect fainter and more extended
emission near the boundary of the Origem Loop than ever before. The
loop consists of four major parts: an elongated arc structure
extending from $\ell = 197\fdg6$ to $\ell = 192\fdg1$ in the north,
which can also be recgonized in the 178~MHz map shown by
\citet{Berkhuijsen74}; the \ion{H}{II} region BFS~52; and two
complexes formed by several known \ion{H}{II} regions e.g. SH 2-261,
SH 2-254 to SH 2-258, the object G192.8$-$1.1 in the south and south
west; and another group of \ion{H}{II} regions SH 2-268, SH 2-270 in
the south east. We marked the names of these known \ion{H}{II} regions
in Fig.~\ref{name}.

\subsection{Spectral indices and their distribution}

\begin{figure}
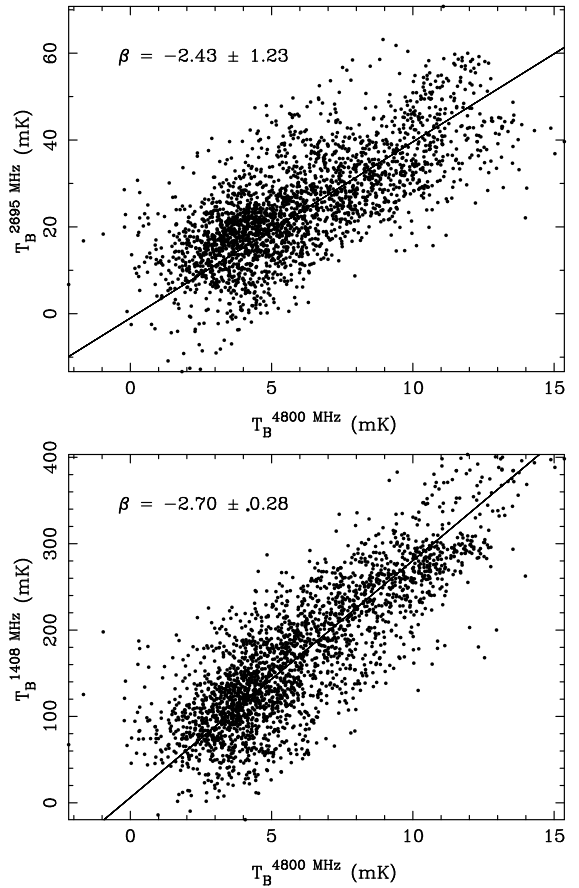
 
\centering
\includegraphics[angle=-90, width=0.4\textwidth]{6_11.ps}
\includegraphics[angle=-90, width=0.4\textwidth]{6_21.ps}
\caption{TT-plot for all the pixels in the Origem Arc region
    (outlined by the black dashed line in Fig.~\ref{name}) between
    $\lambda$6\ cm and $\lambda$11\ cm ({\it upper panel}) and between
    $\lambda$6\ cm and $\lambda$21\ cm ({\it bottom panel}).}
\label{spectrum}
\end{figure}

\begin{figure*}[tbp]
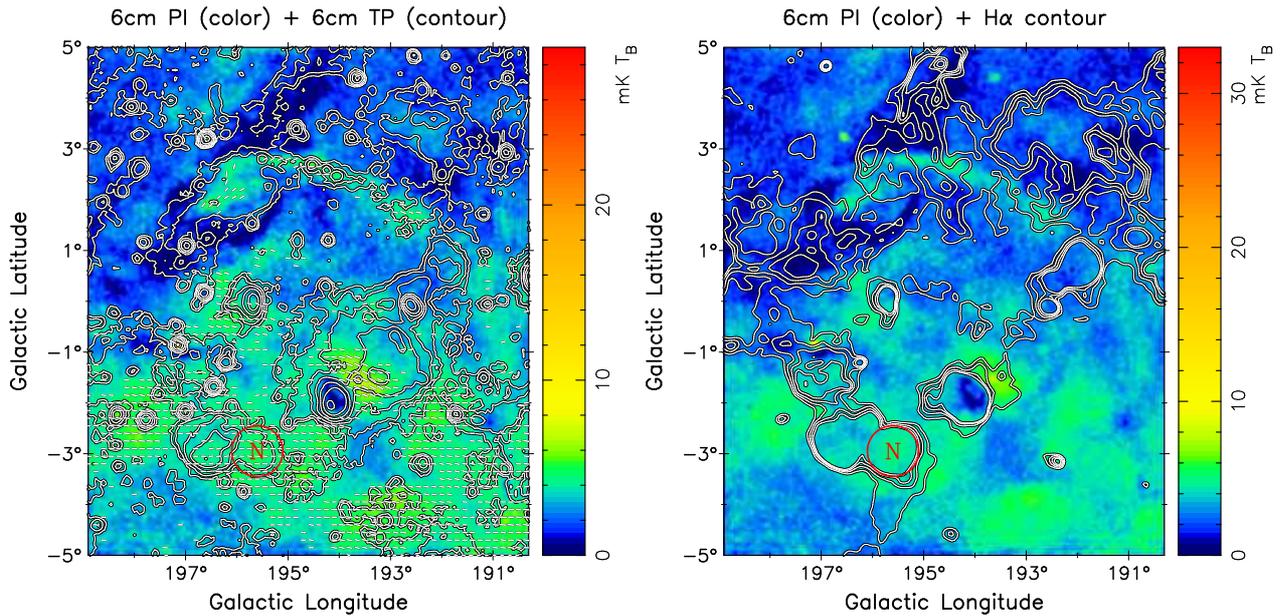

\centering
\includegraphics[angle=-90, width=0.45\textwidth]{Gshell_6pi_color.ps}
\includegraphics[angle=-90, width=0.45\textwidth]{Gshell_6piha_color.ps}
\caption{{\it left panel:} $\lambda$6\ cm polarization image
  (9$\farcm$5 resolution) of the Origem Loop region superimposed with
  total intensity contours and polarization vector $\vec{B} = \vec{E}
  + 90\degr$. {\it right panel:} $\lambda$6\ cm polarization image
  overlaid with the H$\alpha$ intensity (6$\arcmin$ resolution)
  contours. Red circles with ``N'' inside in both images indicate a
  probable new \ion{H}{II} region, see detail in the text.}
\label{6cmpol}
\end{figure*}

Spectral indices and their distribution are important properties for
understanding the nature of the extended radio sources. Shell-type
SNRs usually have a brightness temperature spectral index of $\beta
\sim -2.5$ (T$_{b} \sim \nu^{\beta}$), while the spectrum of an
optically thin \ion{H}{II} region is much flatter, normally being
$\beta \sim -2.1$. We derived the brightness temperature spectral
index distribution of the Origem Loop (see Fig~\ref{Gshell} (d)) using
the $\lambda$6\ cm, $\lambda$11\ cm and the $\lambda$21\ cm images at
an angular resolution of 9$\farcm$5.  A systematic error of such a
spectral index map comes from the uncertainties of the baselevel
determination due to the foreground/background subtraction. In
Fig.~\ref{Gshell} (d), the spectral index map shows reasonable thermal
spectra for all known \ion{H}{II} regions, which demonstrates that the
result is acceptable.  The northern arc of the Origem Loop, which we
call the Origem Arc, obviously has a non-thermal brightness
temperature spectral index around $\beta \sim -2.7$ (flux density
spectral index $\alpha = \beta + 2 = -0.7$). This region was once
singled out by \citet{Krymkin88} in their brightness temperature scans
and designated as GR 0625+16.  They suggested that GR 0625+16 is a
discrete SNR with an integrated radio spectral index of $\alpha =
-0.48\pm0.05$. Although this spectral index is larger than that we
derived from our new data, both indicate the non-thermal nature of the
Origem Arc.

The TT-plot method \citep{Turtle62} was used to verify the spectra
(see Fig.~\ref{spectrum}). The background point sources were
subtracted first, as done in \citet{Gao11y}.  All images were then
smoothed to a common angular resolution of 9$\farcm$5.  For the entire
Origem Arc spanning from $\ell = 197\fdg6$ to $\ell = 192\fdg1$ as
indicated in Fig.~\ref{name}, we obtained the spectral index of
$\beta_{6-11} = -2.43\pm1.23$ from all the data pixels of
$\lambda$6\ cm and $\lambda$11\ cm, and $\beta_{6-21} = -2.70\pm0.28$
for $\lambda$6\ cm and $\lambda$21\ cm. Although the temperature
  measurements for each pixel are not independent, the TT-plots give
  the correct brightness temperature spectral indices and the
  uncertainty estimates, as we tested by using the independent
  pixels. The TT-plot of the brighter part (high signal-to-noise
ratio) of the arc ($\ell \ge 195\fdg0$) gives consistent results as
$\beta_{6-11} = -2.45\pm1.06$, and $\beta_{6-21} = -2.65\pm0.29$. All
these spectral values agree well with the spectral index map shown in
Fig.~\ref{Gshell} (d).

The other parts of the Origem Loop have different properties. The
well-known \ion{H}{II} region, BFS~52 \citep{Blitz82}, has the central
coordinates of $\ell = 191\fdg90, b =0\fdg85$, and the quasar
J061357.6+130645 \citep{Aslan10} is located at $\ell = 197\fdg00, b =
-2\fdg15$. Both have a flat spectrum ($\beta \sim -2.1$). A circular
region centered at $\ell = 195\fdg60, b=-2.95$ with a diameter of
1$\degr$ was found to be very interesting. It has a brightness
temperature spectral index of about $\beta \sim -2.5$ according to
Fig.~\ref{Gshell} (d). TT-plot of this region gives a consistent
result of $\beta_{6-21} = -2.33\pm0.23$, but this also implies a
possibility of being a flat-spectrum thermal emission. We assign its
name G195.60$-$2.95 and marked it using a circle with a central ``N''
in Fig.~\ref{name} and 4. It has strong H$\alpha$ emission and
ring-shaped dust emission (see Fig.~\ref{6cmpol} and 7). The large
ratio between the 60$\mu$m infrared and the $\lambda$6\ cm continuum
emission ($\sim$ 1400) indicates it as a probable thermal \ion{H}{II}
region. G192.8$-$1.1 has a flat thermal spectrum and is not a SNR, as
discussed in \citet{Gao11x}.

\subsection{Polarization}

Observations of polarized emission at $\lambda$6\ cm, $\lambda$1.3\ cm
and DRAO $\lambda$21\ cm were available for the Origem Loop
region. However, only the $\lambda$6\ cm data shows the weak polarized
emission associated with the Origem Arc (see the {\it left panel} of
Fig.~\ref{6cmpol}), in addition to the diffuse polarized background
emission in the lower part of the map. At $\lambda$6\ cm, the
polarized emission is clearly detected within the arc even at the
western end where the total intensity becomes very weak. The
polarization fraction is about 40\% on average. The polarization
B-field vectors ($\vec{E} + 90\degr$) are found to follow the arc,
indicating the presence of tangential magnetic fields. We also noticed
that the depolarization zones seen at $\lambda$6\ cm are correlated
with the enhanced H$\alpha$ emission (see the {\it right panel} of
Fig.~\ref{6cmpol}), e.g. the area around $\ell \sim 192\fdg0, b \sim
3\fdg0$, probably due to the Faraday rotation caused by the magnetic
fields and the thermal electrons. A shuttle-shaped depolarization zone
is found to cross the Origem Arc from northwest to southeast, where
bright H$\alpha$ filaments have good positional correspondences and
morphological similarities.
 
In the southern part of the Origem Loop region, a few large
polarization patches were detected within and outside the
loop. However, none of them seems to be related either to the Origem
Loop or G192.8$-$1.1 \citep{Gao11x}. No arc-shaped structure in
polarization can be found. In the area of $\ell = 194\fdg10, b =
-1.85$, \ion{H}{II} region SH 2-261 acts as a Faraday screen.

At $\lambda$1.3\ cm, no polarized emission is visible in the entire
area of the Origem Loop. Using the average brightness temperature of
the polarized emission in the arc at $\lambda$6\ cm, 5.0~mK\ T$_{b}$,
and the spectral index of $\beta =-2.70$, we estimated the brightness
temperature of the polarized emission of the arc at $\lambda$1.3\ cm,
to be about 0.07~mK T$_{b}$ . It is about the same level of the noise
in the K-band data, which could account for the non-detection of
polarization.

At $\lambda$21\ cm, no correlated polarized emission was detected in
the Origem Arc from the DRAO data, neither. The beam size of the DRAO
data is 36$\arcmin$. Beam and depth depolarization could diminish any
polarized emission. The non-detection might also imply a very near
polarization horizon at $\lambda$21\ cm in this direction.

\subsection{Distances of the Origem Arc and \ion{H}{II} regions}

\begin{table*}
\centering
\renewcommand{\arraystretch}{1.2}
   \caption{Distance of the known \ion{H}{II} regions. The source
     names are listed in the column (1) and the central coordinates
     are in the columns (2) and (3). The radial velocities of the peak
     CO emission associated with the \ion{H}{II} regions are given in
     the column (4). The columns (5) and (6) are the distances in
     \citet[][B74]{Berkhuijsen74}, and \citet[][C85]{Caswell85}. The
     distances from the recent literature are listed in the column (7)
     with the distance measurement method given in the column (8), and
     the references are given in the column (9).}
   \label{t2}
  {\begin{tabular}{ccccccccc}\hline\hline

\multicolumn{1}{c}{Name} &\multicolumn{1}{c}{$\ell$}    &\multicolumn{1}{c}{$b$}       &\multicolumn{1}{c}{CO velocity}       &\multicolumn{1}{c}{Distance (B74)}    &\multicolumn{1}{c}{Distance (C85)}    &\multicolumn{1}{c}{Recent distance}    &Method      &\multicolumn{1}{c}{References} \\
                         &\multicolumn{1}{c}{($\degr$)} &\multicolumn{1}{c}{($\degr$)} &\multicolumn{1}{c}{(km/s)} &\multicolumn{1}{c}{(kpc)}       &\multicolumn{1}{c}{(kpc)}    &\multicolumn{1}{c}{(kpc)}      &\multicolumn{1}{c}{see notes}  &\multicolumn{1}{c}{} \\
\multicolumn{1}{c}{(1)}  &\multicolumn{1}{c}{(2)}       &\multicolumn{1}{c}{(3)}       &\multicolumn{1}{c}{(4)}    &\multicolumn{1}{c}{(5)}         &\multicolumn{1}{c}{(6)}      &\multicolumn{1}{c}{(7)}        &\multicolumn{1}{c}{(8)}        &\multicolumn{1}{c}{(9)}\\
\hline
 BFS 52        &191$\fdg$90    &+0$\fdg$85       &7.3$\pm$0.5         &$\cdots$        &$\cdots$       &$2.10_{-0.026}^{+0.027}$/$1.59_{-0.06}^{+0.07}$ &p$^{\ast}$ &1/2 \\ 
 SH 2-253      &192$\fdg$23    &+3$\fdg$59       &14.4$\pm$0.5        &$\cdots$        &$\cdots$       &5.1$\pm$1.5              &s   &3 \\
 SH 2-254      &192$\fdg$49    &-0$\fdg$15       &7.5$\pm$0.7         &1.12$\pm$0.92   &2.5            &$1.59_{-0.06}^{+0.07}$      &p$\dagger$   &2 \\
 SH 2-255      &192$\fdg$64    &-0$\fdg$01       &7.5$\pm$0.7         &0.88$\pm$0.80   &2.5            &$1.59_{-0.06}^{+0.07}$      &p   &2 \\
 SH 2-256      &192$\fdg$62    &-0$\fdg$13       &7.5$\pm$0.7         &$\cdots$        &2.5            &$1.59_{-0.06}^{+0.07}$      &p$\dagger$   &2 \\
 SH 2-257      &192$\fdg$61    &-0$\fdg$07       &7.5$\pm$0.7         &1.34$\pm$0.95   &$\cdots$       &$1.59_{-0.06}^{+0.07}$      &p$\dagger$   &2 \\
 SH 2-258      &192$\fdg$73    &+0$\fdg$05       &7.5$\pm$0.7         &$\cdots$        &2.5            &$1.59_{-0.06}^{+0.07}$      &p$\dagger$   &2 \\
 SH 2-259      &192$\fdg$94    &-0$\fdg$58       &22.8$\pm$0.5        &$\cdots$        &8.3            &8.9$\pm$2.7              &s   &3 \\
 SH 2-261      &194$\fdg$10    &-1$\fdg$90       &$\cdots$            &0.90$\pm$0.80   &2.3/2.1        &1.6$\pm$0.5              &s   &3 \\
 SH 2-266      &195$\fdg$66    &-0$\fdg$08       &31.2$\pm$1.1        &$\cdots$        &$\cdots$       &2.0$\pm$0.6              &s   &3 \\
 SH 2-267      &196$\fdg$20    &-1$\fdg$20       &$\cdots$            &$\cdots$        &$\cdots$       &$\sim$4.2                &s   &4 \\
 SH 2-268      &196$\fdg$40    &-2$\fdg$80       &4.8$\pm$0.5         &$\cdots$        &$\cdots$       &1.3$\pm$0.4              &s   &3 \\
 SH 2-269      &196$\fdg$40    &-1$\fdg$70       &17.5$\pm$0.7        &1.88$\pm$0.80   &$\cdots$       &$5.28_{-0.22}^{+0.24}$      &p   &5 \\
 SH 2-270      &196$\fdg$84    &-3$\fdg$11       &25.6$\pm$0.4        &$\cdots$        &$\cdots$       &6.8$\pm$2.3              &k   &3 \\
 SH 2-271      &197$\fdg$77    &-2$\fdg$31       &20.5$\pm$0.5        &$\cdots$        &$\cdots$       &5.1$\pm$1.1              &s   &3 \\
 SH 2-272      &197$\fdg$81    &-2$\fdg$28       &20.5$\pm$0.5        &$\cdots$        &$\cdots$       &5.1$\pm$1.1              &s   &3 \\
\hline
\multicolumn{9}{l}{{\bf Notes:} p: parallax; p$^{\ast}$: associated with SH 2-252 or SH 2-254;}\\
\multicolumn{9}{l}{\hspace{8.5mm} p$\dagger$: SH 2-254 to 258 are regarded to have the same distance;}\\
\multicolumn{9}{l}{\hspace{8.5mm} s: stellar distance; k: kinematic distance.}\\
\end{tabular}}
\vspace{-1mm}
\tablebib{
(1) \citet{Reid09}, (2) \citet{Rygl10},  (3) \citet{Russeil03}, (4) \citet{Lahulla87}, (5) \citet{Honma07}}
\vspace{-1mm}
\end{table*}

The observed tangential magnetic fields within the Origem Arc and the
non-thermal spectrum are key evidence for its identification as a
SNR. To verify if this SNR and the \ion{H}{II} regions located in the
southern part of the Origem Loop are physically related, we need to
know their distances first.

Despite of the large uncertainty, the empirical relation between
surface brightness and diameter ($\Sigma$-D) of SNRs provides distance
estimates of shell-type SNRs in case that no related \ion{H}{I} or
molecular clouds (MC) are associated with the SNR. For the entire
Origem Arc, a sector with an opening angle of 128$\degr$, the flux
density at $\lambda$6\ cm is measured to be S$_{6cm}$ = 8.5$\pm$0.9~Jy
after subtracting the background sources. We extrapolate S$_{6cm}$ to
the flux density at 1~GHz with the spectral index of $\alpha =
-0.7$. Using the arc radius of 3$\fdg$3 measured on the map, we
obtained a radio surface brightness of the Origem Arc of $\Sigma_{\rm
  1~GHz}$ = 8.6$\pm0.9\times10^{-23} \rm
Wm^{-2}Hz^{-1}sr^{-1}$. According to the $\Sigma$-D relation found by
\citet{Case98}, the diameter of the Origem Loop is estimated to be
about 195~pc and the distance to be about 1.7~kpc. Note, however, that
as emphasized by \citet{Case98}, the deviation of an individual
estimate derived from their work can be as large as 40\% , which puts
the Origem Arc within a distance range between 1.0 and
2.4~kpc. \citet{Asvarov06} built a model which describes the evolution
of the surface brightness and the diameter of shell-type SNRs with
time.  From his $\Sigma$-D relation (see his Fig.~6), we estimated the
Origem Arc to have a diameter between 100 and 300~pc, corresponding to
a distance range of 0.9 to 2.6~kpc. We therefore conclude that the
Origem Arc is likely to have a distance of 1.7$\pm$0.8~kpc.

The average kinematic distance of several \ion{H}{II} regions located
in the southern part of the Origem Loop was found to be
1.2$\pm$0.7~kpc \citep{Berkhuijsen74}.  However, as noted by
\citet{Caswell85}, the kinematic distances have large
uncertainties. More accurate distance determinations towards these
\ion{H}{II} regions have been obtained through the photometric and
trigonometric parallaxe measurements as listed Table~\ref{t2}, except
for SH 2-270. The lower half of the Origem Loop consists of the
\ion{H}{II} regions BFS~52, SH 2-254 to 258, SH 2-261, SH 2-268 and
the \ion{H}{II} region SH 2-266, which have similar distances as to
the Origem Arc, and the other distant \ion{H}{II} regions SH 2-253, SH
2-267, SH 2-269, SH 2-271 and SH 2-272. We could not get the distances
of the object G192.8$-$1.1 and the newly identified object
G195.60$-$2.95.

\subsection{Siginitures at other wavelengths}

\citet{Berkhuijsen74} searched for possible \ion{H}{I} structures
associated with the Origem Loop, however, no clue was found.
\citet{DeNoyer77} proposed that an \ion{H}{I} jet, which appears at
$\ell \sim 197\degr, b \sim 2\degr$ may be related to the Origem Arc,
although the jet is extended beyond its boundary. This jet is a part
of the prominent \ion{H}{I} structure, the Anti-centre shell (ACS),
which was discovered by \citet{Heiles84}. We checked the new
\ion{H}{I} data from the Leiden/Argentina/Bonn \ion{H}{I} survey
\citep{Hartmann97, Kalberla05} and the GALFA \ion{H}{I} DR1
data\footnote{https://purcell.ssl.berkeley.edu/index.php} for a much
larger area (20$\degr \times 20\degr$). We found that the ACS is
prominent in the negative velocity map and disappears in the positive
velocity map. This clearly differs from the positive CO radial
velecity associated with the \ion{H}{II} regions discussed
above. Moreover, the ACS has a much larger size ($\sim$ 30$\degr$ in
diameter) than the Origem Loop. For the velocity 0.0 to 32.5~km/s, no
associated \ion{H}{I} structure is found around the Origem Loop.

\begin{figure} 
\centering
\includegraphics[width=0.45\textwidth]{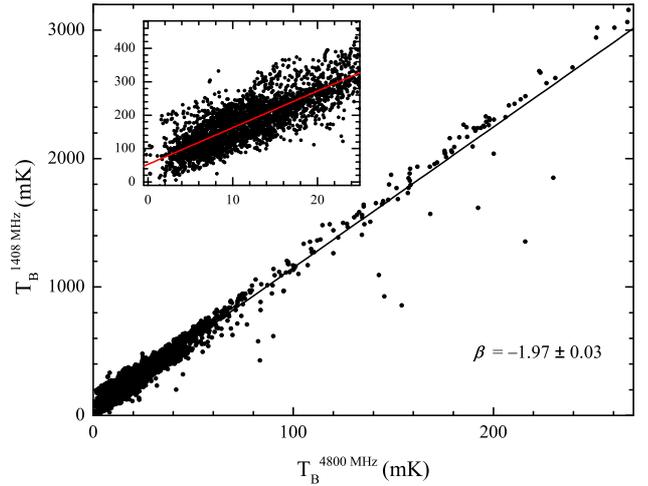}
\caption{TT-plot for the southern part of the Origem Loop between
  $\lambda$6\ cm and $\lambda$21\ cm. The inner small image is the
  zoom-in picture for the value T$_{6cm}$ $\leqslant$
  25~mK\ T$_{b}$. The red line in the small image represents the same
  spectral index as in the large image.}
\label{TT2}
\end{figure}

The TT-plot can be used to reveal different emitting components with
different spectral indices \citep[e.g.][]{Xiao08}. The TT-plot of the
southern half of the Origem Loop shows that thermal emission is
overwhelmingly dominant. Unlike in the northern arc, no evidence was
found for any detectable unambiguous non-thermal emission component in
the south (see Fig.~\ref{TT2}).

\citet{Berkhuijsen74} investigated the relation between the Origem
Loop and the \ion{H}{II} regions. Based on the age of the loop and the
evolution timecale from protostars to the \ion{H}{II} regions, it was
hard to tell whether the Origem Loop triggered the star formaion that
leads to the \ion{H}{II} regions in the south. Using the physical size
determined in this work, adopting equation (3) in
\citet{Berkhuijsen74}, the Origem Arc is about 1 to 3~Myr
old. \citet{Chava08} estimated the ages of the \ion{H}{II} regions SH
2-254 to 258 from an expansion model of a Str{\"o}mgren sphere,
ranging between 0.1~Myr to 5.0~Myr. \citet{Bieging09} proposed a
sequential star formation scenario from interaction between the
\ion{H}{II} regions and molecular clouds which indicates that at least
the younger \ion{H}{II} regions SH 2-255 to 257 were triggered by SH
2-254.

\begin{figure*}
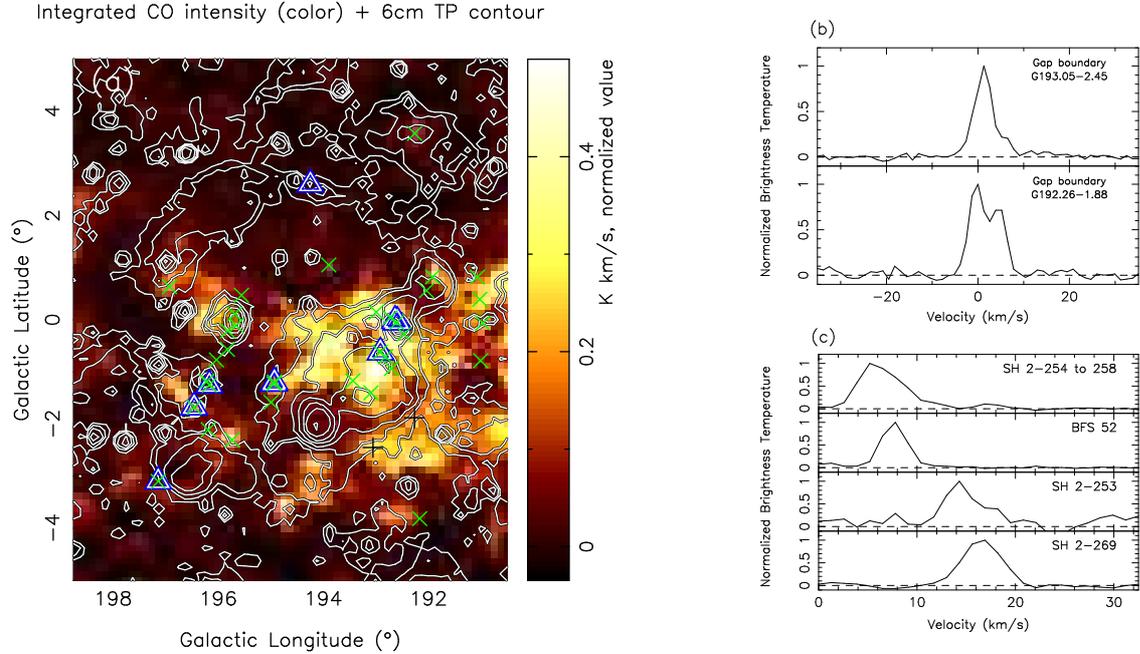
 
\centering
\begin{minipage}[bth]{0.45\textwidth}
\centering
\includegraphics[angle=-90,width=8.0cm]{COCO.cps}
\end{minipage}
\begin{minipage}[bth]{0.45\textwidth}
\centering
\includegraphics[angle=-90,width=5.0cm]{compCO_boundary.ps}\\
\includegraphics[angle=-90,width=5.0cm]{compCO.ps}
\end{minipage}
\caption{{\it Left panel}: CO intensity map integrated from 0.0 to
  32.5~km/s, overlaid by the $\lambda$6\ cm total intensity contours
  as shown in Fig.~\ref{Gshell}. Each pixel value was normalized by
  deviding by the maximum integrated intensity in the image. The green
  crosses represent the proto-stellar candidates selected from the
  IRAS point source catalog by the criteria introduced by
  \citet{Junkes92} while the blue-white triangles are the massive YSOs
  found in the Red MSX survey. {\it Top right panel:} Normalized CO
  emission profile for the two areas marked with the black ``plus'' in
  the {\it left panel}. {\it Bottom right panel}: CO spectra for the
  \ion{H}{II} regions SH 2-254 to 258, BFS~52, SH 2-253, and SH
  2-269. The brightness temperature in each region was also
  normalized.}
\label{CO}
\end{figure*}

The interaction between the SNR shock and the ambient molecular clouds
should broaden the linewidth. For example, IC~443, a famous SNR-MC
interaction case, clearly shows a high CO~J=3$-$2/CO~J=2$-$1 ratio and
the broadening of CO emission lines \citep{Xu11}. Broadened CO~J=1$-$0
lines were also detected in another SNR-MC interaction case
\citep{Byun06}. Public CO~J=3$-$2 and CO~J=2$-$1 data covering the
Origem Loop region are not available.  We checked the CO~J=1$-$0 data
of the Origem Loop region from the CO
survey\footnote{http://www.cfa.harvard.edu/rtdc/CO/} \citep{Dame01}.
The radial velocities of the CO emission peaks associated with these
\ion{H}{II} regions were previously measured and given by
\citet{Blitz82}, as we listed in our Table~\ref{t2}. Therefore, we
integrated the CO emission in the velocity range from 0.0~km/s to
32.5~km/s and show the result in the {\it left panel} of
Fig.~\ref{CO}. We find that the most intense CO emission comes from
the area of G192.8$-$1.1. It has a roughly similar morphology as the
$\lambda$6\ cm continuum emission as illustrated by the contour lines
in Fig.~\ref{CO}. An obvious gap of CO emission is seen at the edge of
the continuum emission in the southwest of G192.8$-$1.1. We checked
the velocity-intensity relation at two positions indicated by ``plus''
in the left panel of Fig.~\ref{CO} and also the velocity-intensity
plots for the \ion{H}{II} region SH 2-254 to SH 2-258, BFS~52 for the
possible interaction group and the \ion{H}{II} regions SH 2-253, SH
2-269 for the non-interaction group. As shown in the {\it right panel}
of Fig.~\ref{CO}, all of the radial velocities corresponding to the
peak CO emission for the \ion{H}{II} regions are consistent with those
found by \citet{Blitz82}. We checked the linewidths for several
hundred non-interacting \ion{H}{II} regions \citep{Anderson09,
  Russeil04} and found the average values are in the range of 3 $\sim$
6~km/s. SH 2-254 to 258, BFS~52, SH 2-253 and SH 2-269 have similar
linewidths and no significant rising of the line wings can be
seen. Therefore there are no hints for the interaction between SNR and
clouds for \ion{H}{II} region formation.

We searched for massive young stellar objects in the Origem Loop
region from the Red MSX Source Survey
database\footnote{http://www.ast.leeds.ac.uk/RMS/} and for the
protostellar candidates in the IRAS point source catalog. They are
marked in the Fig.~\ref{CO}. Several of them are coincident and most
of them are located in the regions where CO emission is prominent. The
young stellar objects in the Origem Loop region do not show over
density neither on the northern Origem Arc nor on the shell-like
structure where the continuum-CO boundary exists.

\begin{figure} 
\centering 
\includegraphics[angle=-90,width=0.45\textwidth]{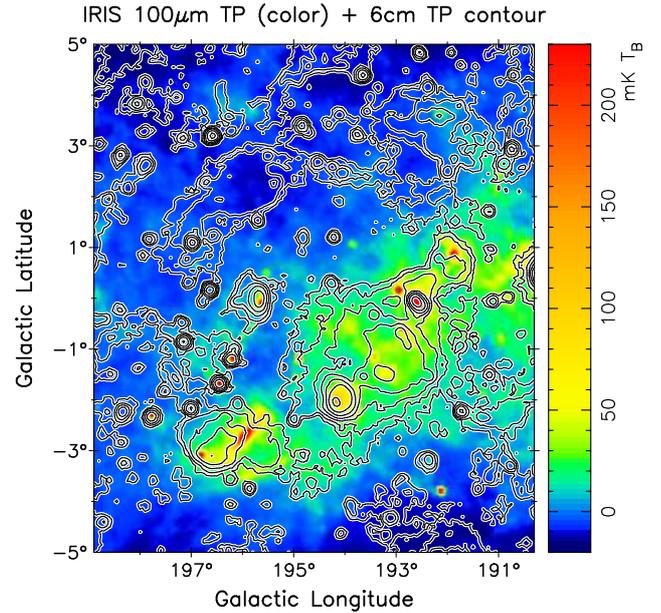} 
\caption{100$\mu$m dust map (2$\arcmin$ resolution) overlaid with the
  contours of the $\lambda$6\ cm total intensity. The contours are the
  same as in Fig.~\ref{Gshell}.}
\label{infrared}
\end{figure}

Infrared (dust) image was also checked by \citet{Berkhuijsen74},
nothing coincided with the Origem Arc (see her Fig.~3). From the IRIS
100$\mu$m dust image \citep{Miville05} of the Origem Loop region shown
in Fig.~\ref{infrared}, we found that the dust emission is well
correlated with the \ion{H}{II} regions in the south. The two
interacting \ion{H}{II} regions SH 2-255 and SH 2-257 triggered star
formation in between them \citep{Ojha11}. A large number of YSOs were
identified on their boundary. The object G195.6-2.95 has an incomplete
ring structure, and the infrared emission gets enhanced between it and
the known \ion{H}{II} regions SH 2-268 which is about 1.3~kpc
away. However, we do not get any YSOs on the intensive infrared
emission zone. It might be due to limitation of the infrared data sets
we use. The apparent sizes of the infrared bubbles in the inner Galaxy
are generally smaller than the one around G195.6$-$2.95
\citep{Simpson12}, which may imply a small distance to
G195.6$-$2.95. An infrared loop was also found around the SNR
\citep{Koo08}. However, the polarization measurement at $\lambda$6\ cm
and the infrared/radio ratio of G195.6-2.95 strongly suggest that it
is an \ion{H}{II} region.

We checked the 0.1-0.4~keV and 0.4-2.4~keV ROSAT X-ray
images\footnote{http://www.xray.mpe.mpg.de/cgi$_{-}$bin/rosat/rosat$_{-}$survey},
no associated structure with the Origem Loop was found.

\section{Summary}

We used multi-frequency survey data to revisit the Origem Loop in the
anticentre of the Galaxy. The Origem Arc is a polarized non-thermal
emission structure with a spectral index of $\beta = -2.70$,
indicating that it is a shell-type SNR. We estimated its distance to
be 1.7$\pm$0.8~kpc.

Using the new radio data, we discussed the possibilities of a physical
association between the SNR and the \ion{H}{II} regions located in the
south of the Origem Loop. Inspection of TT-plots for different
emission components, the width of CO lines and age estimates did not
give evidence of a non-thermal southern arc or the interaction between
the SNR and the \ion{H}{II} regions in the south. Associated infrared
emission is seen to be well related to the \ion{H}{II} regions in the
southern part of the loop. No \ion{H}{I} or X-ray emission correlated
with the Origem Loop was found.

\begin{acknowledgements}
We would like to thank the referee, Dr. Elly Berkhuijsen for
constructive and helpful comments, which significantly improve the
paper.  The Sino-German $\lambda$6\ cm polarization survey was carried
out with a receiver system constructed by Mr. Otmar Lochner at MPIfR
mounted at the Nanshan 25-m telescope of the Urumqi Observatory of
NAOC. The MPG and the NAOC/CAS supported the construction of the
receiving system by special funds.  We thank Mr. Maozheng Chen and
Mr. Jun Ma for qualified maintenance of the receiving system for many
years.
The authors are supported by the National Natural Science foundation
of China (10833003) and the Partner group of the MPIfR at NAOC in the
frame of the exchange program between MPG and CAS for many bilateral
visits. XYG is additionally supported by the Young Researcher Grant of
National Astronomical Observatories, Chinese Academy of Sciences.
This paper made use of information from the Red MSX Source survey
database at www.ast.leeds.ac.uk/RMS which was constructed with support
from the Science and Technology Facilities Council of the UK. This
paper also utilizes data from Galactic ALFA HI (GALFA HI) survey data
set obtained with the Arecibo L-band Feed Array (ALFA) on the Arecibo
305m telescope.  Arecibo Observatory is part of the National Astronomy
and Ionosphere Center, which is operated by Cornell University under
Cooperative Agreement with the U.S. National Science Foundation. The
GALFA HI surveys are funded by the NSF through grants to Columbia
University, the University of Wisconsin, and the University of
California.

\end{acknowledgements}

\bibliographystyle{aa}
\bibliography{bbfile}

\end{document}